# Optical Activity at 1.55 μm in Si:Er:O Deposited Films

S. A. Abedrabbo [1,2 (a)] and A. T. Fiory [2]

1. Department of Physics, University of Jordan, Amman, Jordan 11942.
2. Department of Physics, New Jersey Institute of Technology, Newark, N.J. 07901.

**Abstract**

Silicon films doped with Er and O were prepared by techniques of physical vapor deposition on crystalline silicon, ion beam mixing and oxygen incorporation through $Ar^+$ and $O_2^+$ implantation, and thermal annealing. Processing steps were tailored to be compatible with standard CMOS and to be of notably low cost to fabricate optically active media for silicon-based infrared emitters. The Si:Er:O films exhibit strong photoluminescence at room temperature that is analogous to Stark-split $Er^{+3}$-ion 1.55-μm bands in fiber-optic materials. Concentration distributions were determined by Rutherford backscattering spectrometry. It is found that photoluminescence signals increase with the O to Er ratio. Ion implantation effectively enhances the thermal diffusion of Er and improves its optical activation.

## 1. Introduction

For electronic and photonic integrated circuits based on all-silicon optoelectronics embedded in CMOS (complementary metal oxide semiconductor) technology, the silicon-based light source remains a particularly uncertain technology component. To become the dominant light source, it is of major importance to develop room temperature efficient silicon light emitters. This is a sensible direction in light of the enormous silicon investment deployed worldwide, which stood at 63 billion US dollars in 2007 (EETIMES, 2008). The obstacle remains that intrinsic Si is an inferior light emitter owing to its indirect band gap. Among the many efforts taken to overcome this barrier, impurity center luminescence by incorporation of rare-earth atoms stands out as a promising technique (Ebans-Freeman, 2001). Attention is centered on erbium because its radiative de-excitation at 1.5 μm matches a low absorption band in silica fibers. For this reason Er has been and remains the choice for optical network amplifiers, such as in erbium-doped fiber amplifiers (EDFA) and erbium-doped waveguide amplifiers (EDWA) (Yam, 2006; Ramamurthy,1998; Thomson, 2005); Thomson, 2006). Consequently, it is sensible to pursue light

---

[(a)] Email: sxa0215@yahoo.com; sxa0215@njit.edu.

emitting diode (LED) and related optoelectronic devices that incorporate this same impurity center.

Two interrelated issues that one may face when incorporating Er in Si are the low solubility of Er in Si, estimated to be $1 \times 10^{18}$ cm$^{-3}$ (Coffa, 1998), and its tendency to segregate by precipitation when approaching maximum permissible solubility concentrations (Eaglesham, 1991). To date, numerous methods have been explored to incorporate Er and other rare-earth elements in Si with the aim of improving its light emission properties. The most promising techniques generally utilize processes that are compatible with the microelectronics industry and yield high concentrations of optically active Er in Si. Some of these processes that had success in circumventing the solubility limitations are ion implantation and solid phase epitaxy (Polman, 1993; Custer, 1994; Liu, , 1995), and film deposition by molecular beam epitaxy (Ni, 1997; Stimmer, 1996) or chemical vapor deposition (Morse, 1996). Moreover, it is necessary to also consider processes that circumvent thermal quenching, which is a fundamental physical property of Er impurity centers in crystalline silicon that otherwise greatly attenuates their optical activity at room temperature (Scalese, 2000).

A promising approach to resolve the above issues has been to introduce various O-rich Er structures in Si (Coffa, 1998; Castagna, 2006; Priolo, ,1993), given that Er$^{+3}$ is optically active in silica (e.g., as utilized for EDFAs and EDWAs). In a study of p$^+$/n$^+$ junctions, where the n$^+$ electrode was doped by implantation of Er and O, luminescence was observed under reverse bias (hot carrier excitation). In another study an Er-rich dielectric layer was formed within a metal oxide semiconductor (MOS) device, yielding 0.2% external optical efficiency at room temperature with reasonable device stability (Castagna, 2006). Weiss et al. have demonstrated a tunable light source in the near infrared that consists of porous silicon microcavities infiltrated with erbium doped nematic liquid crystals (Weiss, 2007). Whereas the Er is the luminescing source, the porous microcavities narrow the emitted band while the liquid crystal enables wavelength tuning.

Previous work (Abedrabbo, 2009) demonstrated the feasibility of creating room-temperature optically active Er$^{3+}$ in a semiconductor-based film by a relatively straightforward process involving physical vapor deposition (PVD) of elemental Si and Er and thermal annealing. Oxygen was introduced by taking advantage of the oxygen-gettering affinity of Er metal as well as by oxygen-ion implantation. The study included ion-beam mixing through Ar$^+$ implantation. Optically active Er$^{+3}$ in the resulting Si:Er:O structures was demonstrated by room temperature photoluminescence. This work presents a new analysis of the previously prepared samples in terms of understanding the relative importance of the two main points introduced above, namely the roles of the Er concentration and the presence of oxygen as a co-species in forming optically active Er$^{+3}$ at room temperature. Section 2 briefly reviews the experimental procedures and Section 3 presents the experimental results. Section 4 discusses the interpretation of the results according to the line of inquiry stated above. Section 5 summarizes the results, presents conclusions, and discusses their implications for process improvement in silicon-based optoelectronics.

## 2. Experimental Procedure

The objective of the sample preparation technique for this work is to produce films comprising silicon, one or more rare earths, and an activating dopant, principally oxygen. In the fabrication of Si:Er:O films, the dopant is erbium and the activator is oxygen. The initial step in the process is physical vapor deposition of a Si-Er alloy by vacuum co-evaporation from two separately controllable resistively heated boats with each containing an elemental source. The substrates are 1cm$^2$ silicon wafer samples cleaved from double-side polished wafers of p-type doping (10$^{15}$ cm$^{-3}$ boron) and are held at 300 °C. One evaporation boat is charged with Si cut from the substrate material, while the other is charged with Er metal in a crucible. The evaporation takes place within a bell jar chamber at 10$^{-4}$ Pa. Native oxides on the substrates and charges in the evaporation boats were retained prior to film deposition. Films with two Si-Er compositional profiles, corresponding to Er concentrations from 20 to 50 %, and two thicknesses (nominally 200 and 300 nm) were prepared. Film thicknesses were estimated from in-situ readings of a quartz-crystal mass-thickness monitor (subsequently determined post-processing independently, as discussed below).

Two samples were also implanted with Ar$^+$ (300 keV energy, 10$^{18}$ cm$^{-2}$ dose, depth range 170 to 410 nm) to induce ion-beam mixing and with O$_2^+$ (260 keV, 5×10$^{17}$ cm$^{-2}$, depth range 140 to 410 nm) for oxygen incorporation. The implantations were performed in the Jordan Van de Graaff accelerator (JOVAC) facility available at the University of Jordan. Oxygen was implanted since it is known to enhance the optical activity of Er$^{+3}$ centers in silicon (Coffa,1998; Priolo, 1993). The motivation for the Ar$^+$ implants follows from previous work, which showed that ion-beam mixing leads to nonequilibrium or metastable-alloy films on surfaces (Abedrabbo, 2006). By randomizing the Er site symmetry in this manner, one anticipates hindering Er precipitation, thereby addressing the issue of exceeding solid solubility levels.

The final process step for all the samples is annealing in vacuum (600 °C, 1 hr.). Table I lists the four samples studied in this work. Silicon-erbium films for samples 1 and 2 were deposited simultaneously (i.e., samples 1 and 2 received the same Si and Er PVD) and similarly for samples 3 and 4. Samples 1 and 3 were annealed after growth. Samples 2 and 4 received both the Ar$^+$ and the O$_2^+$ ion implants prior to annealing. The sample matrix thus comprises two as-deposited film thicknesses and either with or without the dual ion implants. Implantation leads to increased final film thickness, as discussed in Sect. 3.

Rutherford backscattering spectrometry (RBS) and model fits were used to obtain concentration depth profiles of the constituent elements, Si, Er, O, and Ar. For RBS analysis, an energetic 2-MeV He$^+$ beam at normal incidence to the sample is utilized, with the detector (Si-based) collecting scattered He$^+$ ions at 122°. Raw RBS spectra of the samples were reported previously (Abedrabbo, 2009; Mohammed, 2007). The software package SIMNRA (Simulation of Nuclear Reaction Analysis, offered by the Max Planck Institute) was utilized to simulate the spectra by representing the samples as multilayer structures (equivalent to 24 nm per layer on average) comprising Si, Er, O, and Ar (where implanted) on a Si substrate. Best fits of the simulated spectra to the RBS data were obtained by varying as adjustable parameters the fractions of the constituent species within the model layers. The results of this analysis yield atomic concentrations of the constituent species as functions of depth. From the species distributions one obtains Si:Er:O film thickness and average atomic concentrations of Er and O, denoted as $<c_{Er}>$ and $<c_O>$, respectively, which are given in Table I.

TABLE I. Thickness (from RBS analysis), implant specifications (details in text), integrated photoluminescence (PL) intensity, and average atomic concentrations of erbium and oxygen, $<c_{Er}>$ and $<c_O>$, respectively for the Si:Er:O samples processed in this work.

| Sample | Si:Er:O Thickness (nm) | Implants | Integrated PL Intensity (a.u.) | $<c_{Er}>$ (at. %) | $<c_O>$ (at. %) |
|---|---|---|---|---|---|
| 1 | 268 | (none) | 126 | 15 | 29 |
| 2 | 524 | O$_2^+$, Ar$^+$ | 188 | 5 | 14 |
| 3 | 199 | (none) | 65 | 22 | 13 |
| 4 | 285 | O$_2^+$, Ar$^+$ | 65 | 12 | 13 |

Room temperature photoluminescence (PL) was used to evaluate infrared emission from the $Er^{3+}$ centers utilizing a model Flourolog-3 spectrofluorometer. A Xe lamp fixed at 530 nm by a double excitation monochromator was used for excitation while a Hamamatsu InGaAs photodiode detector, preceded by a single emission monochromator, determined signal intensity. Although the PL measurements were done after the RBS, effects from the He beam are likely to be inconsequential, partly because RBS probed only about 10% of the film area measured by PL (estimated from beam spot size ratio), but mainly because He is a light mass (relative to Er) and inefficient in producing ion-beam damage at the low dosages of RBS analysis ($\sim 3 \times 10^{15}$ cm$^{-2}$).

## 3. Experimental Results

Photoluminescence spectra verify that this PVD-based technique for Si:Er:O film preparation leads to the formation of optically active $Er^{3+}$ at room temperature in the four samples studied. Spectra for samples 1 and 2 are shown in Fig. 1. While the Si:Er deposition was the same for both samples, sample 1 was unimplanted and sample 2 received the $Ar^+$ and $O_2^+$ implants. Spectra such as observed in all samples (Abedrabbo, 2009) are typical of the superimposed manifold of $Er^{+3}$ Stark-split $^4I_{13/2}$ - $^4I_{15/2}$ transitions for erbium dopants in solid hosts and observed at room temperature in the presence of inhomogeneous broadening (Desurvire, 1990; Desurvire, 1994). Dominant wavelengths and relative strengths of the transitions were determined by modeling PL spectra as a superposition of up to five wavelengths each with Gaussian broadening (Abedrabbo, 2009). The two strongest transitions are found to be in the vicinity of 1516 and 1535 nm, which correspond to energy level transitions from 6644 $cm^{-1}$ to 51 $cm^{-1}$, and from 6644 $cm^{-1}$ to 125 $cm^{-1}$, respectively. Although the relative strengths of the transitions are known to depend on local fields and structural environments at the $Er^{+3}$ ions (e.g., $Er-O_6$ octahedral co-ordination in silica), a general theoretical model is formidable challenge and for room temperature is presently unavailable. This is the case especially when random host structure, leading to various site symmetries, permits maximum number of transitions (Desurvire, 1994). Generally, the studied PL spectra show small sample-to-sample variations in mean wavelength (1516 to 1520 nm) and width, as determined from the root-second moment (18 to 24 nm).

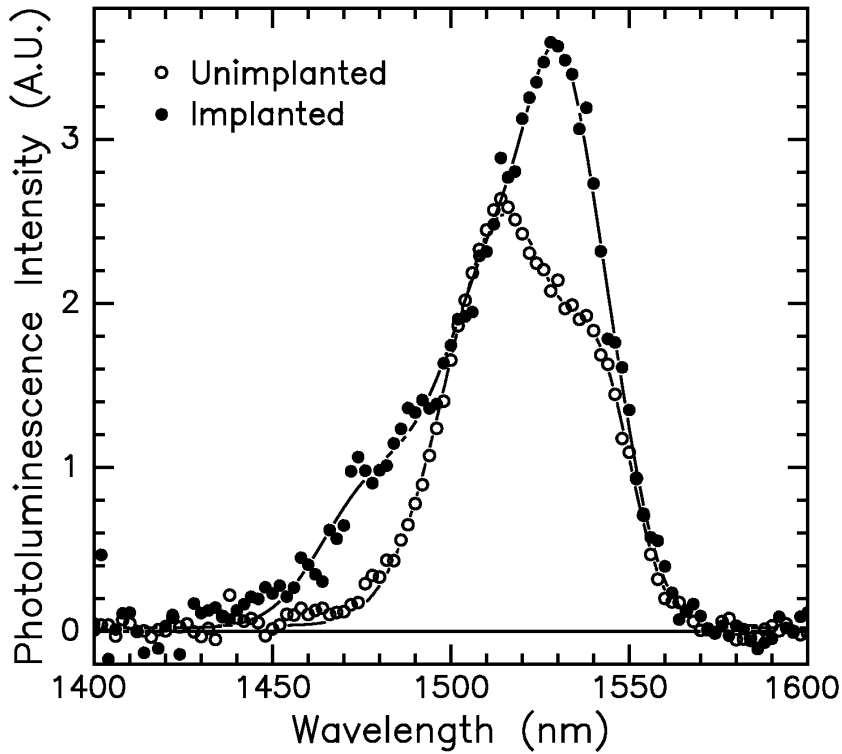

Figure 1. Room-temperature photoluminescence spectra of Si:Er:O films annealed at 600 °C. Open symbols – sample 1, unimplanted; closed symbols – sample 2, implanted with $Ar^+$ and $O_2^+$.

Integrated PL intensity, defined as area under the PL spectra (determined by numerical integration), provides a useful measure for comparing relative strengths of $Er^{3+}$ optical activity among various samples. Results of this analysis are given in Table I. Integrated PL intensity

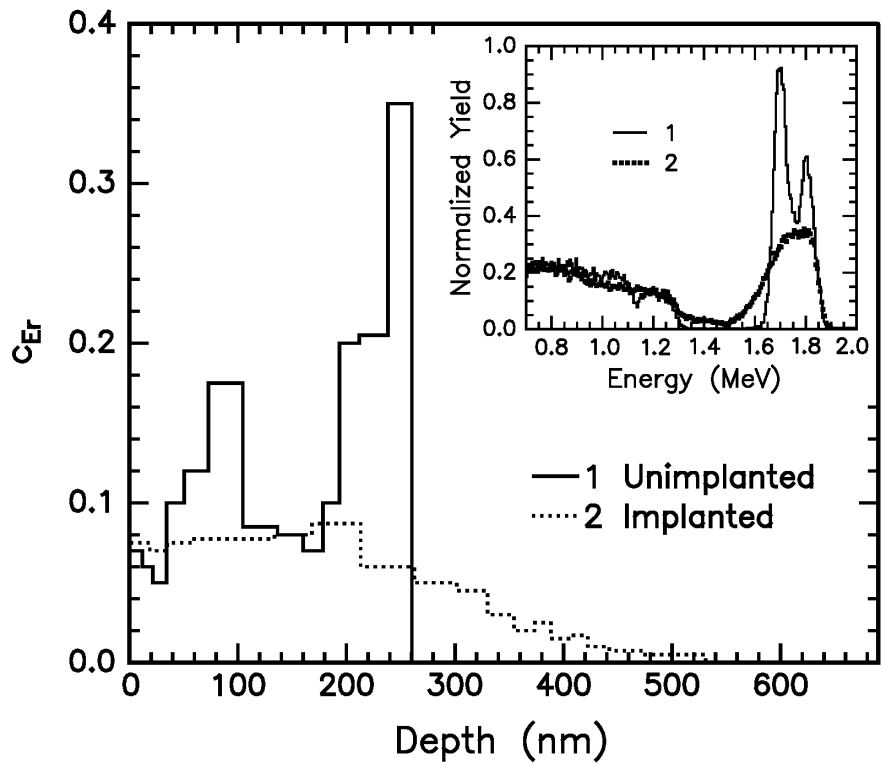

Figure 2. Depth variation of erbium atomic concentration, $c_{Er}$, in annealed Si:Er:O films: sample 1, unimplanted (solid) and sample 2, implanted with $Ar^+$ and $O_2^+$ (dotted). Inset: normalized RBS spectra for samples 1 and 2.

appears to be only weakly correlated with film thickness, if at all.

The increase in $Er^{3+}$ optical activity with implantation is associated with a redistribution of the Er concentration. The Er atomic fractional concentration $c_{Er}$ (a dimensionless quantity), obtained by fitting RBS with layered structures, is shown in Fig. 2 as a function of depth for samples 1 and 2. The inset to Fig.2 shows the normalized raw RBS spectra. The peaks between 1.6 and 1.9 MeV correspond to backscattering from Er. Range of backscattering energy scales approximately with depth as 0.96 keV/nm. The maximum backscattering energies (in MeV) for the various species are: Er, 1.88; Ar, 1.44; Si, 1.27; and O, 0.90. From this analysis one finds that ion implantation of sample 2 creates a broadened Er distribution and a lower average Er concentration (see Table I), when compared to unimplanted sample 1.

RBS analysis also reveals the presence of oxygen in all samples, whether implanted with oxygen or not (atomic oxygen concentrations are given in Table I). The ubiquitous presence of oxygen in all samples indicates that the scavenging effect of the highly-reactive Er plays a role in oxygen incorporation during film processing as well as from oxidation of surface Er-Si structures.

Figure 3 (a) reveals a major finding of this investigation. Here, the integrated PL intensity is plotted against the ratio of the average oxygen and erbium atomic concentrations $\langle c_O \rangle / \langle c_{Er} \rangle$ (equivalently, $N_O/N_{Er}$, where $N_O$ and $N_{Er}$ are numbers of oxygen and erbium atoms, respectively, per unit area in the Si:Er:O films, as determined by RBS). Points in this and subsequent figures are labeled with sample numbers in correspondence with Table I. The dashed curve (serving as guide to the eye) highlights a clear correlation: the PL intensity increases with O/Er atomic ratio.

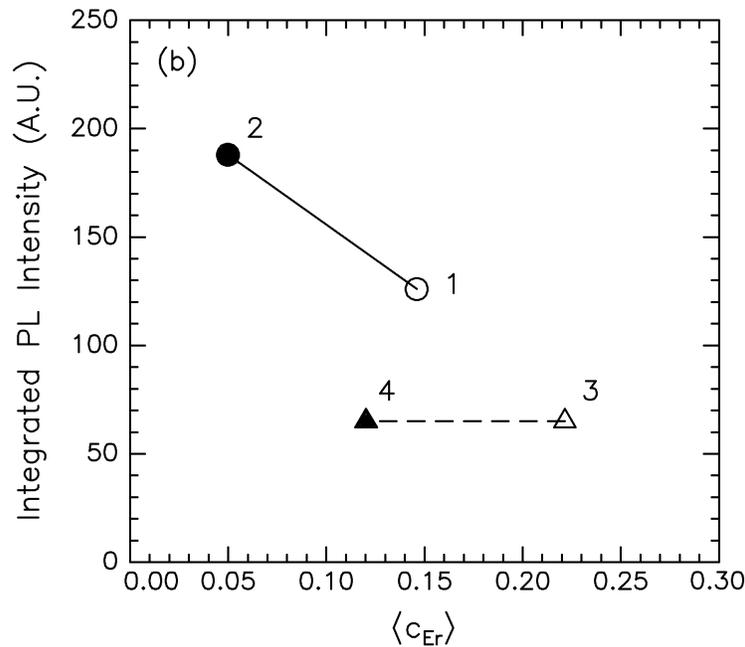

Figure 3 (a). Variation of integrated photoluminescence with ratio of atomic concentrations of oxygen and erbium, $\langle c_O \rangle / \langle c_{Er} \rangle$. Labels denote sample numbers, Table I. Dashed curve is guide to the eye.

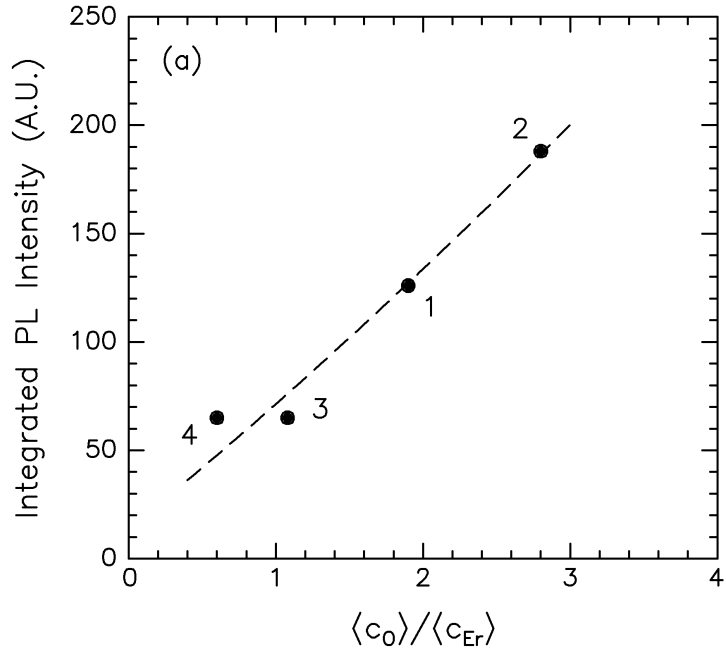

Figure 3 (b). Variation of integrated photoluminescence with average Er atomic concentration, $\langle c_{Er}\rangle$, for unimplanted samples 1 and 3 (open symbols) and implanted samples 2 and 4 (filled symbols). Lines connect samples with the same as-deposited Si:Er film.

Figure 3 (b) shows the variation of PL intensity with average atomic concentration of Er, $\langle c_{Er}\rangle$, displayed as distinct symbols for each sample. Lines connect points for related samples (i.e., with identically same PVD process). Solid points denote unimplanted samples and filled points denote implanted samples. The data show a tendency (i.e., evidence for correlation) for PL intensity to decrease with increasing $\langle c_{Er}\rangle$. Inspection of Table I shows weak correlation also with thickness of Si:Er:O film. Thus the data of Fig. 3 show that the optically active $Er^{3+}$ systematically increases with $\langle c_O\rangle/\langle c_{Er}\rangle$ while it decreases with $\langle c_{Er}\rangle$. The explanation for this behavior is that optical activity of $Er^{3+}$ depends on the formation of Er-O ligands (Coffa, 1998), the density of which increases with $\langle c_O\rangle/\langle c_{Er}\rangle$. The decrease in photoluminescence signal with increasing $\langle c_{Er}\rangle$ suggests that the number of Er-O ligands per Er is suppressed at high Er concentration. From the statistics of random mixtures, one expects this result based on bonding competition with neighboring erbium and silicon atoms.

The effects of ion implantation on Er concentration profiles in four samples are summarized in Fig. 4, where the Er area atomic density $N_{Er}$ is plotted against the average Er atomic concentration $\langle c_{Er}\rangle$. Samples 1 and 3, which are denoted by open symbols, were deposited without ion implantation. Sample 3 (open triangle symbol) contains the greatest as-deposited atomic concentration of Er. For the ion implanted samples 2 and 4 (closed symbols) the average Er atomic concentrations are reduced relative to the respective unimplanted samples 1 and 3. These findings are the result of two effects: ion-beam damage leads to enhanced diffusion of the Er (see Fig. 2) and reduces the average atomic concentration, $\langle c_{Er}\rangle$, while sputtering of the films reduces the area concentration $N_{Er}$. Denoting Si:Er:O film thickness as $d_i$, where i = 1, 2, 3 and 4 for the four samples, the data in Table I reveal that $(d_2 - d_1)$ exceeds $(d_4 - d_3)$ by nearly a factor 3. One may understand this as greater efficiency of inducing ion beam mixing in the thicker film of sample 2 ($d_2$ = 268 nm), because more of the $Ar^+$ implant (range 170 to 410 nm) stops in that film. Hence, the implanted-$O^+$ in sample 2 fell in proximity to ion-beam-diffused Er forming Er-O and enhancing Er optical activity, while implanted-$O^+$ in sample 4 was concentrated at farther depths saturating defect traps in Ar-amorphised Si (Polman, 2001) and not enhancing Er optical activity.

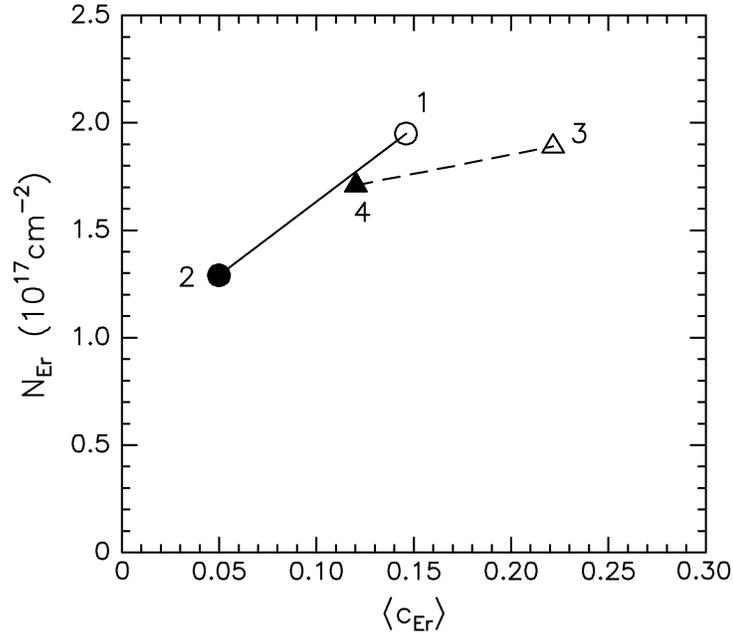

Figure 4. Variation of Er area concentration, $N_{Er}$, with average Er atomic concentration, $\langle c_{Er} \rangle$, for unimplanted samples 1 and 3 (open symbols) and implanted samples 2 and 4 (filled symbols). Lines connect samples with the same as-deposited Si:Er film.

In summary, the primary experimental findings are (1) PL signals increase with the ratio of O to Er concentrations (Fig 3a), showing that optically active $Er^{3+}$ in Si:Er:O structures occurs for Er in association with O; (2) PL signals tend to decrease with increasing Er concentration $\langle c_{Er} \rangle$, which was studied over the range 5 to 22 % (Fig 3b). These concentration dependences provide an insight as to why ion implantation improves the PL signal only for sample 2. One notes that the ion implanted sample 2 has a higher ratio $\langle c_O \rangle / \langle c_{Er} \rangle$, which is a consequence of a broader Er distribution (Fig. 2) that substantially decreases the $\langle c_{Er} \rangle$ value in the denominator. Moreover, the number of oxygen atoms in the Si:Er:O films of samples 1 and 2 (e.g., determined either from area concentration $N_O$ or as product of film thickness times $\langle c_O \rangle$, see Table I) are almost the same (6% less O in sample 2). Similar oxygen content suggests that Er-O scavenging reactions are dominant mechanisms for incorporating oxygen atoms.

## 4. Discussion

Although the formation of Er-O and Er-O-Si complexes are known to contribute to the optical activity of $Er^{+3}$ ions, a precise mechanism for light output enhancement due to the presence of O has not been clearly established (Michel, 1998). One possibility is that the concentration of Er in Si may increase in presence of O impurities due to the formation of localized Er-O complexes, as opposed to optically inactive Er precipitation. Another is that an Er-O ligand complex is formed that provides dipole coupling to transfer electron-hole recombination energy to the 4f Er manifold. A third postulate is that ligand ions enhance the solubility of Er in Si by the very formation of Er-O localized complexes (Michel, 1998). The ligand essentially provides the necessary energy-transfer from silicon band-to-band transitions to Er impurities in order to cause the inversion and excitation needed for photon emission. High electronegativity elements like O are well known to increase the optical activity of Er by means that are not restricted to increasing solid solubility or decreasing Er segregation. When introduced into the solid host, oxygen increases the absorption probability and therefore the emission yield (Anderson, 1996).

Ion implantation, which is utilized here in a post-deposition treatment, plays an important role in redistributing the Er and enhancing its association with oxygen. Atomic concentration profiles for samples 1 and 2 shown in Fig. 2 demonstrate this effect. Ion beam mixing leads to a broadened Er distribution as a consequence of irradiation-induced diffusion of the Er that appears in the subsequent annealing step. In this case enhanced diffusion also smears out the nonuniformity in the as-deposited film. Diffusion enhancement varies with the overlap of the implanted-ion and deposited-Er profiles. Enhanced diffusion extends the depth of the Er distribution on the order of 100 nm, as evident by comparing the Er distribution profiles shown in Fig. 2. This enhanced diffusion is defined in terms of the difference in the variance of the Er distribution for the irradiated sample 2 to that of the un-irradiated sample 1 (Abedrabbo, 2009). These results show that the Er redistribution is assisted by Er-O reactions that are promoted by ion-beam damage in the amorphous deposited film as well as in the crystalline Si substrate.

## 5. Summary and Conclusions

The principle findings of this work are that optical activity of $Er^{+3}$ increases with the O/Er atomic concentration ratio (Fig. 3a), with the highest PL signal obtained for $<c_O>/<c_{Er}> = 2.8$ (sample 2). Data trends (Fig. 3b) suggest a tendency for PL signals to decrease with increasing $<c_{Er}>$. One therefore concludes that Er-O association promotes formation of optically active $Er^{+3}$, while Er-Er association tends to inhibit it. The strongest PL signal corresponds to $<c_{Er}> = 5\%$, which although at the lower end of the studied range, is already a rather high Er concentration, when compared to previous studies [typically less than 0.2% (Scalese, 2000)]. Thus the samples under study seem to be close to optimizing the tradeoffs between the desirable Er-O and undesirable Er-Er associations. One could take these results into consideration for improving the PVD process for manufacturing LEDs. For example, the O/Er ratio could be increased by alternative means, such as by varying film deposition conditions (e.g., increasing film thickness and using slightly lower Er concentration) so as to lower the doses for ion-implantation and ion-beam mixing.

The present results portend a promising approach for LED applications. The combination of physical vapor deposition, ion implantation, and thermal annealing is notable for its ease in reaching high concentrations of Er (including the silicide). Although the importance of oxygen availability on light emission at room temperature from silicon doped with a rare-earth metal has been demonstrated, optimal compositions for applications remain to be determined. Further study of the nature of energy transfer among the various rare-earth metals may lead to an overall improvement in emission efficiency. This includes radiative levels of multi-doped silicon matrices and the overall silicon optical properties in the IR spectral range as functions of doping with a variety of rare-earth metals.

The film deposition approach is ideally suited to forming LED structures of the p-n junction type where the n-type Si side is rare-earth doped, e.g., with Er. It is also suited to forming Schottky diodes with integrated $ErSi_{1.7}$ contacts (Wu, 1996; Jang, 2003). In such applications, the temperature and thermal budget may be adjusted to anneal out crystalline damage produced by the ion-beam mixing process. The depth of the mixed layers of silicon, rare earths, and oxygen, which control the width of the depletion region and overall performance of LED structures, would be tailored to optimum photon emission efficiency.

## Acknowledgments

This work was supported by the assistance of the Deanship of Academic Research at the University of Jordan, Project contract no. 1030 and Hamdi Mango Center for Scientific Research (HMCSR) and the New Jersey Institute of Technology.

The authors are indebted to Q. Mohamed for his assistance during his research assistantship at the University of Jordan, and to Professor N.M. Ravindra of New Jersey Institute of Technology for his continuous interest, support and encouragement. We would like to thank the JOVAC accelerator crew and Professor D.E. Arafah, Horiba Jobin-Yvon Fluorescence Division in Edison, New Jersey, and the Physics Department at Rutgers University, New Brunswick, New Jersey, for their invaluable assistance. Publication on this work has appeared (Abedrabbo 2010).